\documentstyle[prl,aps,epsf,calc,graphicx]{revtex}

\begin{document}

\title{Josephson Effect Between Triplet Superconductors:\\
A Self-Test for the Order Parameter Symmetry of Bechgaard Salts}

\author{C. D. Vaccarella, R. D. Duncan and C. A. R. S\'a de Melo} 
\address{School of Physics, 
Georgia Institute of Technology, Atlanta, GA 30332} 

\date{\today}
\maketitle

\begin{abstract}
We show that the Josephson
effect between triplet superconductors is sensitive to the relative
orientation of the ${\bf d}$-vectors across the junction. In addition,
we point out that the temperature and angular 
dependence of the Josephson effect
can help distinguish between different order parameter symmetries.
We discuss coherent and incoherent tunneling processes,
and the role of surface bound states
within the framework of non-local lattice Bogoliubov-deGennes equations. 
\vspace{1pc}
\end{abstract}


Recent upper critical field and NMR experiments in 
${\rm (TMTST)_2 X}$~\cite{lee-97,lee-98,lee-02}, 
where ${\rm X} = {\rm ClO_4},~{\rm PF_6}$ have indicated that 
they may be unconventional triplet superconductors, while low temperature
thermal conductivity experiments~\cite{belin-97} suggest 
that the superconducting state is gapped.
These experiments were inspired by 
early suggestions of triplet superconductivity
in Bechgaard salts~\cite{abrikosov-83,lebed-86,dms-93,sdm-96}, 
and served as inspirations for the intense theoretical 
efforts~\cite{lebed-99,sdm-99,lebed-00,shimahara-00,kuroki-01,duncan-01}
that followed.
Lebed, Machida, and Ozaki (LMO)~\cite{lebed-00} 
proposed a ``p-wave'' triplet order parameter for 
${\rm (TMTSF)_2 PF_6}$,
where the ${\bf d}$-vector had a strong
component along the ${\bf a}$ direction, thus producing a strongly anisotropic spin susceptibility 
with $\chi_{a} \ll \chi_N$ and $\chi_{b^\prime} \approx \chi_N$.
A fully gapped singlet ``d-wave'' order parameter
for ${\rm (TMTSF)_2 ClO_4}$ was 
proposed by Shimahara~\cite{shimahara-00},
while gapless triplet ``f-wave'' superconductivity 
for ${\rm (TMTSF)_2 PF_6}$ was proposed by 
Kuroki, Arita, and Aoki (KAA)~\cite{kuroki-01}.
Duncan, Vaccarella and S\'a de Melo (DVS)~\cite{duncan-01}
performed a detailed group theoretical analysis and 
suggested that a weak spin-orbit fully gapped 
triplet ``$p_x$-wave'' order parameter, where
$\chi_{a} \approx \chi_N$ and 
$\chi_{b^\prime} \approx \chi_N$~\cite{sdm-99,sdm-98},
would be a good candidate for superconductivity 
in Bechgaard salts.

Here, we propose a test for 
the order parameter symmetry of Bechgaard salts
which is based on the Josephson effect 
between two triplet superconductors (TS) separated by 
an insulating (I) barrier.
Our suggestion differs
from previous proposals invoking tests of the symmetry of the order
parameter in the context of triplet heavy fermion
superconductors~\cite{pals-77,larkin-86},
which relied on the Josephson effect between a {\it singlet} 
and a {\it triplet} superconductor. 
We show that the Josephson 
effect between two TS exists in the simplest case
where the tunneling matrix element is spin-conserving 
(with or without time-reversal invariance), 
and that it depends on the relative orientation of the
${\bf d}$-vectors of the TS. 
In addition, we show that the temperature and angular 
dependence of the 
Josephson effect can also help distinguish between different triplet
states. 
We include coherent and incoherent tunneling processes and
the effects of surface bound states (within the framework of
non-local lattice Bogoliubov-deGennes equations).


In anticipation of the existence of surface bound 
states in particular geometries of 
unconventional superconductors~\cite{buchholtz-81,hu-94},
we prefer to use a real space representation of the 
{\it left} $(L)$ and {\it right} $(R)$ superconductors
Hamiltonian
\begin{equation}
\label{eqn:L-R-hamiltonian}
H_j = 
\sum_{ {\bf r}_j, {\bf r^\prime}_j } 
{\cal H} ({\bf r}_j, {\bf r^\prime}_j) 
c_{j,\alpha_j}^{\dagger} ({\bf r}_j) 
c_{j, \alpha_j} ({\bf r^\prime}_j)
+ 
\sum_{ {\bf r}_j, {\bf r^\prime}_j } 
\left[
\Delta_{j,\alpha_j, \beta_j} ({\bf r}_j, {\bf r^\prime}_j)
c_{j,\alpha_j}^{\dagger} ({\bf r}_j) 
c_{j, \beta_j}^{\dagger} ({\bf r^\prime}_j)
+ H. C.
\right],
\end{equation}
where indices $\alpha_{j}$ and $\beta_{j}$ (with $j = $ $L$,$R$)
are spin labels, ${\bf r}_j, {\bf r^\prime}_j$ are position labels. 
Repeated greek indices indicate summation.
The first term of $H_j$ contains  ${\cal H} ({\bf r}_j, {\bf r^\prime}_j) 
= \left[ t_j ({\bf r}_j, {\bf r^\prime}_j) - 
\mu \delta_{{\bf r}_j, {\bf r^\prime}_j} \right]$, where
$t_j ({\bf r}_j, {\bf r^\prime}_j)$ are transfer integrals confined
to nearest neighbors and $\mu_j$ are the chemical potential.
The second term of $H_j$ contains the order parameter matrix
$
\Delta_{j,\alpha_j, \beta_j} ({\bf r}_j, {\bf r^\prime}_j)
= 
V_{j, \alpha_j, \beta_j, \delta_j, \gamma_j} ({\bf r}_j, {\bf r^\prime}_j)
\langle  
c_{j, \delta_j} ({\bf r^\prime}_j)
c_{j,\gamma_j} ({\bf r}_j) 
\rangle,
$
where 
$V_{j, \alpha_j, \beta_j, \delta_j, \gamma_j} ({\bf r}_j, {\bf r^\prime}_j)$
is the two-body pairing interaction.
The tunneling Hamiltonian connecting $L$ and $R$ 
superconductors is
\begin{equation}
\label{eqn:tunneling-hamiltonian}
H_T = 
\sum_{ {\bf r}_L, {\bf r}_R }
\left[
T_{\alpha_L \alpha_R} ({\bf r}_L, {\bf r}_R )
c^\dagger_{L,\alpha_L} ({\bf r}_L) c_{R,\alpha_R} ({\bf r}_R)
+
H. C. 
\right],
\end{equation}
where the matrix element for tunneling is
$
T_{\alpha_L \alpha_R} ({\bf r}_L, {\bf r}_R)
= T_s ({\bf r}_L, {\bf r}_R) \delta_{\alpha_L, \alpha_R}
+ {\bf T}_v ({\bf r}_L, {\bf r}_R) \cdot
\sigma_{\alpha_L, \alpha_R},
$
with the first term $T_s$ corresponding to spin-conserving tunnelling, 
and the second term ${\bf T}_v$ corresponding to spin-dependent tunnelling.
The pair current through the junction 
is (to second order in the tunneling matrix element)
\begin{equation}
\label{eqn:pair-current}
J_s (V, T) = 
{
2e
\over 
\hbar
}
{\it Im} 
\left[ \exp (i\omega_J t)
\sum_{ {\bf r}_L {\bf r}_R \atop {\bf r^\prime}_L {\bf r^\prime}_R}
T_{\alpha_L \alpha_R}({\bf r}_L, {\bf r}_R)
T_{\beta_L, \beta_R}({\bf r^\prime}_L, {\bf r^\prime}_R)
P_{\alpha_R \beta_R}^{\alpha_L \beta_L} 
({\bf r}_L, {\bf r}_R, {\bf r^\prime}_L, {\bf r^\prime}_R, i\omega_n) 
\right],
\end{equation}
where $i\omega_n$ corresponds to Matsubara frequencies
continued to $(-eV/\hbar + i\delta)$, and 
$\omega_J = 2eV/\hbar$ with $V$ being 
the voltage applied across the junction.
The tensor $P_{\alpha_R \beta_R}^{\alpha_L \beta_L}$
can be written in terms of the anomalous Green's matrices as
$
P_{\alpha_R \beta_R}^{\alpha_L \beta_L} 
({\bf r}_L, {\bf r}_R, {\bf r^\prime}_L, {\bf r^\prime}_R, i\omega_n)
=
T \sum_{i\nu_m} 
F^\dagger_{L,\alpha_L\beta_L} ({\bf r}_L, {\bf r^\prime}_L, i\nu_m)
F_{R,\alpha_R \beta_R} ({\bf r}_R, {\bf r^\prime}_R, i\omega_n - i\nu_m),
$
where 
$
F_{j, \alpha_j \beta_j} ({\bf r}_j, {\bf r^\prime}_j, i\nu_m) 
= T \int d\tau \exp(i\nu_m \tau)
F_{j, \alpha_j \beta_j} ({\bf r}_j, {\bf r^\prime}_j, \tau), 
$
with
$
F_{j, \alpha_j \beta_j} ({\bf r}_j, {\bf r^\prime}_j, \tau), 
=
\langle 
T c_{j,\alpha_j} ({\bf r}_j, \tau) 
c_{j,\beta_j} ({\bf r^\prime}_j, 0) 
\rangle.
$
The general form of the order parameter 
matrix 
$
\Delta_{j,\alpha_j, \beta_j} ({\bf r}_j, {\bf r^\prime}_j) =
i \Delta_{si,j} ({\bf r}_j, {\bf r^\prime}_j) 
\left[ \sigma_2 \right]_{\alpha_j,\beta_j} 
+ i {\bf d}_j ({\bf r}_j, {\bf r^\prime}_j) \cdot 
\left[
\sigma \sigma_2 
\right]_{\alpha_j \beta_j},
$
where the first term corresponds to the 
singlet (pseudo-singlet) state and the second to 
the triplet (pseudo-triplet) state
in the case of weak spin-orbit coupling 
(in the case of strong spin-orbit coupling).
The order 
parameter matrix $\Delta_{j,\alpha_j, \beta_j} ({\bf r}_j, {\bf r^\prime}_j)$
satisfies the Pauli exclusion principle, since 
the function $\Delta_{si,j} ({\bf r}_j, {\bf r^\prime}_j)$ 
which represents singlet pairing is symmetric under
the exchange ${\bf r}_j \leftrightarrow {\bf r^\prime}_j$, 
and the vector ${\bf d}_j ({\bf r}_j, {\bf r^\prime}_j)$ 
which represent triplet pairing is antisymmetric
under the exchange ${\bf r}_j \leftrightarrow {\bf r^\prime}_j$. 
The reduced Hamiltonians on both sides of the junction
can be separately diagonalized via the lattice Bogoliubov-de Gennes
transformation
$
c_{j,{\alpha}_j} ({\bf r}_j) = 
\sum_{N_j} 
\left[ 
u_{N_j \alpha_j} ({\bf r}_j) \gamma_{N_j}  + 
v^*_{N_j \alpha_j} ({\bf r}_j) \gamma_{N_j}^\dagger
\right]
$,
where $u_{N_j \alpha_j}$ and $v_{N_j \alpha_j}$ are two component spinors
for each value of $\alpha_j$ and 
satisfy the corresponding non-local Bogoliubov-deGennes (BdG) equation
\begin{equation}
\label{eqn:bdg1}
E_{N_j} u_{N_j \alpha_j} ({\bf r}_j) = 
\sum_{ {\bf r^\prime}_j } 
{\cal H} ({\bf r}_j, {\bf r^\prime}_j) 
u_{N_j \alpha_j} ({\bf r^\prime}_j)
+ 
\sum_{ {\bf r^\prime}_j } 
\Delta_{j,\alpha_j, \beta_j} ({\bf r}_j, {\bf r^\prime}_j) 
v_{N_j \beta_j} ({\bf r^\prime}_j)
\end{equation}
\begin{equation}
\label{eqn:bdg2}
-E_{N_j} v_{N_j \alpha_j} ({\bf r}_j) = 
\sum_{ {\bf r^\prime}_j } 
{\cal H}^* ({\bf r}_j, {\bf r^\prime}_j) 
v_{N_j \alpha_j} ({\bf r^\prime}_j)
+ 
\sum_{ {\bf r^\prime}_j } 
\Delta^*_{j,\alpha_j, \beta_j} ({\bf r}_j, {\bf r^\prime}_j) 
u_{N_j \beta_j} ({\bf r^\prime}_j).
\end{equation}
This set of equations must be solved self-consistently
together with the order parameter equation
\begin{equation}
\label{eqn:order-parameter}
\Delta_{j,\alpha_j, \beta_j} ({\bf r}_j, {\bf r^\prime}_j) 
= {1 \over 4} 
V_{j, \alpha_j, \beta_j, \delta_j, \gamma_j} ({\bf r}_j, {\bf r^\prime}_j) 
\sum_{N_j} (1 - 2f_{N_j})
\left[ 
v^*_{N_j \gamma_j} ({\bf r}_j) u_{N_j \delta_j} ({\bf r^\prime}_j) 
-
v^*_{N_j \delta_j} ({\bf r}_j) u_{N_j \gamma_j} ({\bf r^\prime}_j) 
\right],
\end{equation}
where the two-body potential is written under the assumption 
of weak spin-orbit coupling as
$
V_{j, \alpha_j, \beta_j, \delta_j, \gamma_j} ({\bf r}_j, {\bf r^\prime}_j) 
=
2 I_{1,j} ({\bf r}_j, {\bf r^\prime}_j) 
\delta_{\alpha_j \gamma_j} \delta_{\beta_j, \delta_j}
+ 2 I_{2,j} ({\bf r}_j, {\bf r^\prime}_j) 
\sigma_{\alpha_j \gamma_j} \cdot \sigma_{\beta_j, \delta_j}
$.

From now on we will assume that 
the superconductors on either side of the 
junction are in 
(a) a triplet unitary state (time reversal invariant state in the bulk);
(b) characterized by weak spin-orbit coupling; 
(c) the tunneling matrix element
$T_{\alpha_L \alpha_R} ({\bf r}_j, {\bf r^\prime}_j)$ 
is spin conserving, i. e., the tunnel barrier preserves 
spin, and thus it is not magnetically active;
and 
(d) the 
${\bf d}_j$-vector is weakly locked to the
${\bf c}_j$ axis by the weak spin-orbit coupling, a choice
that is consistent with 
Knight shift experiments 
of Lee {\it et. al.}~\cite{lee-02}.
All these assumptions seem to be applicable
to the Bechgaard salts. We model these quasi-one-dimensional salts 
via the  bulk dispersion
$
\varepsilon_j ({\bf k}_j) = -|t_{x,j}| \cos(k_{x,j} a_j)
-|t_{y,j}| \cos(k_{y,j} b_j) - |t_{z,j}| \cos(k_{z,j} c_j),
$
where 
$|t_{x,j}| \gg  |t_{y,j}| \gg  |t_{z,j}|$,
corresponding to an orthorhombic crystal 
($D_{2h}$ group) with lattice constants
$a,b$ and $c$ along the ${\bf a},{\bf b}$ and ${\bf c}$ 
axis respectively. 
In the $D_{2h}$ point group all representations are one dimensional 
and non-degenerate~\cite{duncan-01}, which means that
the ${\bf d}$-vector in momentum space for unitary triplet
states in the weak spin-orbit coupling limit
is characterized by one of the four states:
(1) $^3 A_{1u} (a)$, 
with  
${\bf d}_j ({\bf k}) = {\hat \eta_j} \Delta_{f_{xyz},j} X_j Y_j Z_j$ 
(``$f_{xyz}$'' state);
(2) $^3 B_{1u} (a)$, 
with  
${\bf d}_j ({\bf k}) = {\hat \eta_j} \Delta_{p_z, j} Z_j$ (``$p_{z}$'' state);
(3) $^3 B_{2u} (a)$, 
with  
${\bf d}_j ({\bf k}) = {\hat \eta_j} \Delta_{p_y, j} Y_j$ (``$p_{y}$'' state);
(4) $^3 B_{3u} (a)$, 
with  
${\bf d}_j ({\bf k}) = {\hat \eta_j} \Delta_{p_x, j} X_j$ (``$p_{x}$'' state).
Since, the Fermi surface touches
the Brillouin zone boundaries the functions 
$X_j$, $Y_j$, and $Z_j$ need to be periodic and can 
be chosen to be $X_j = \sin{(k_{x,j} a_j)}$, 
$Y_j = \sin{(k_{y,j} b_j)}$, 
and $Z_j = \sin{(k_{z,j} c_j)}$.
The unit vector $\hat \eta_j$ defines the direction 
of ${\bf d}_j ({\bf k})$. Next, we discuss the Josephson
effect in two situations, first neglecting, and second
including surface bound state effects.

{\it The Josephson effect neglecting surface bound states:} 
In this approximation only the contribution from scattering
states with well defined momentum is included. 
Under this assumption the BdG amplitudes become 
$
u_{N_j \alpha_j} ({\bf r}_j) = \exp(i {\bf k}_j \cdot {\bf  r}_j)
\tilde u_{{\bf k}_j, n_j} 
/\sqrt{V};
$
$
v_{N_j \alpha_j} ({\bf r}_j) = 
\exp(i {\bf k}_j \cdot{\bf  r}_j)
\tilde v_{{\bf k}_j, n_j} 
/\sqrt{V},
$
where quantum indices $N_j$ are represented by momentum 
${\bf k}_j$ and discrete index $n_j = 1_j, 2_j$. 
In this case, the expression for $J_{s} (V, T)$ 
defined in Eq.~(\ref{eqn:pair-current})
becomes
\begin{equation}
\label{eqn:pair-current-kspace}
J_s (V,T) = 
{
2e
\over 
\hbar
} 
\left[ \cos(\bar\phi) \sum_{{\bf k}_L {\bf k}_R}
Im
\lbrace
W ({\bf k}_L, {\bf k}_R, i\omega_n)
\rbrace
+
\sin(\bar\phi) \sum_{{\bf k}_L {\bf k}_R}
Re
\lbrace
W ({\bf k}_L, {\bf k}_R, i\omega_n)
\rbrace
\right],
\end{equation}
after summations over spin indices $\alpha_j, \beta_j$,
indices $n_j$,
Matsubara frequencies $i\nu_m$, 
and the global U(1) gauge transformation ${\bf d}_j \to 
{\bf \bar d}_j \exp(i\phi_j)$ are performed.
Here, $\bar\phi = \phi_R - \phi_L + \omega_J t$, and
$
W({\bf k}_L, {\bf k}_R, i\omega_n )
= 
2 
{\bar {\bf d}}_{L}^* ({\bf k}_L) 
{\bf \cdot} 
{\bar {\bf d}}_{R} ({\bf k}_R)
Q({\bf k}_L, {\bf k}_R)
{
P ({\bf k}_L, {\bf k}_R, i\omega_n )
},
$
with
$
Q ({\bf k}_L, {\bf k}_R) =
T_{s}({\bf k}_L, {\bf k}_R) 
T_{s}(-{\bf k}_L, -{\bf k}_R),
$
and $i\omega_n \to -eV/\hbar + i\delta$.
Notice that
the dot product 
${\bf d}_{L}^* ({\bf k}_L) \cdot {\bf d}_{R} ({\bf k}_R)$
appears in $J_s (V,T)$, and thus
$J_s (V,T)$ is very sensitive to the relative orientation between
${\bf d}_{L}^* ({\bf k}_L)$ and 
${\bf d}_{R} ({\bf k}_R)$~\cite{millis-88}. 
For instance, if ${\bf d}_{L}^* ({\bf k}_L) 
\perp {\bf d}_{R} ({\bf k}_R)$ then
$J_s (V,T)$ vanishes identically for any $V$ and $T$. 
Furthermore $J_s (V,T)$ (for fixed $V$ and $T$) changes sign
depending if the vectors 
${\bf d}_{L}^* ({\bf k}_L)$ and ${\bf d}_{R} ({\bf k}_R)$
are aligned or anti-aligned. It is this sensitivity to the
relative orientation of the ${\bf d}$-vectors across the junction that
makes the Josephson effect between TS a crucial
test for triplet superconductivity itself.
The function 
$P ({\bf k}_L, {\bf k}_R, i\omega_n ) = 
T \sum_{i\nu_m} 
\left[ (\hbar\nu_m)^2 + E_{{\bf k}_L}^2 \right]^{-1}
\left[ 
(\hbar\omega_n - \hbar \nu_m)^2 + E_{{\bf k}_R}^2
\right]^{-1} 
$
is a sum over 
Matsubara energies $\hbar \nu_m = (2m + 1)\pi k_B T$.   
The excitation energies are
$
E_{{\bf k}_j} = 
\sqrt
{ 
\xi_{{\bf k}_j}^2 + 
|{\bf d}_j ({\bf k}_j )|^2 
}.
$
Notice that
$J_s (V,T)$ also
vanishes identically
if $W ({\bf k}_L, {\bf k}_R, i\omega_n \to -eV/\hbar + i\delta )$
is odd under inversion for ${\bf k}_L \to -{\bf k}_L$
or ${\bf k}_R \to -{\bf k}_R$. Since
$ {\bf d}_j ( {\bf k}_j ) $ is odd, 
$E_{ {\bf k}_j }$ is even under inversion 
${\bf k}_j \to - {\bf k}_j $, and  
$P ({\bf k}_L, {\bf k}_R, i\omega_n \to -eV/\hbar + i\delta )$ 
is even under 
${\bf k}_L \to - {\bf k}_L $ or 
${\bf k}_R \to - {\bf k}_R $, a non-vanishing $J_s (V,T)$
requires $Q ({\bf k}_L, {\bf k}_R)$ to have contributions which
are odd in both ${\bf k}_L$ and ${\bf k}_R$.
Therefore, $J_s (V,T)$
depends crucially on the magnitude and symmetry 
properties of $Q ({\bf k}_L, {\bf k}_R)$ defined above, which contains
the tunneling matrix element $T_s ({\bf k}_L, {\bf k}_R)$. 
Since,
all the $D_{2h}$ point group representations are one dimensional 
and non-degenerate~\cite{duncan-01},
$T_s$ can be expanded as
$
T_s ({\bf k}_L, {\bf k}_R) = 
\sum_{\Gamma_L \Gamma_R} 
T_{\Gamma_L \Gamma_R} ({\bf k}_L, {\bf k}_R) 
\bar\psi^{\Gamma_L} ({\bf k}_L) 
\psi^{\Gamma_R} ({\bf k}_R), 
$
in the case of weak spin-orbit coupling. 
Here $\psi^{\Gamma_j} ({\bf k}_j)$ are the basis functions of
of the $D_{2h}$ (orthorhombic) point group~\cite{duncan-01}.
In the particular 
case of time reversal symmetry $T_s (-{\bf k}_L, -{\bf k}_R) 
= T_s^* ({\bf k}_L, {\bf k}_R)$ and  
$Q({\bf k}_L, {\bf k}_R) = |T_s ({\bf k}_L, {\bf k}_R)|^2 \ge 0 $. 
\begin{figure} [hb!]
\begin{center}
\epsfxsize=8cm
\epsfysize=8cm
\epsffile{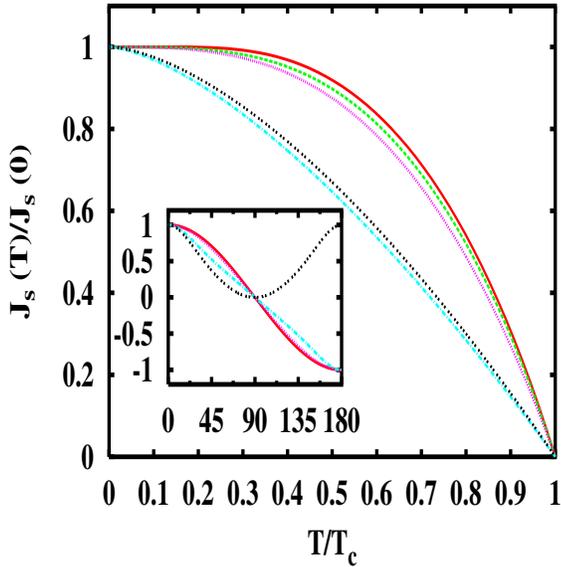}
\vskip 0.3cm
\caption{
Plot of $J_s (T)/J_s (0)$ versus $T/T_c$. Inset shows
$J_s (0,\alpha_{LR})/J_s (0,0)$ versus $\alpha_{LR}$ at 
$T = 0$. Several ${\bf a}$ axis tunneling processes
are illustrated for ``$p_x$'' and ``$p_y$'' symmetries.
{\it Incoherent processes}: ``$p_x$'' with (a) 
$Q^{(inc)} \propto X_L X_R$ (solid line)
and (b) $Q^{(inc)} \propto 1_L 1_R$ (trivially zero);
``$p_y$'' with (c) 
$Q^{(inc)} \propto Y_L Y_R$ (dashed line)
and (d) $Q^{(inc)} \propto 1_L 1_R$ (trivially zero).
{\it Coherent processes}: ``$p_x$'' with (a) 
$Q^{(coh)} \propto X_L X_R$ (dotted line)
and (b) $Q^{(coh)} \propto 1_L 1_R$ (trivially zero);
``$p_y$'' with (c) 
$Q^{(coh)} \propto Y_L Y_R$ (dot-dashed line)
and (d) $Q^{(coh)} \propto 1_L 1_R$ (double-dotted line).
The $L$ and $R$ superconductors are
assumed to be identical quarter-filled systems, with  
$T_{c} = 1.5~{\rm K}$, and parameters
$t_x = 5800~{\rm K}$, $t_y = 1226~{\rm K}$, $t_z = 48{\rm K}$. 
}
\label{fig:one}
\end{center}
\vskip -0.5cm
\end{figure}
For
the weak spin-orbit ``$p_x$'', ``$p_y$'', ``$p_z$''
and ``$f_{xyz}$'' states,
where
the direction of the ${\bf d}$-vectors is independent of 
momentum,
a simple expression for $J_s (T)$ at zero bias can be obtained 
\begin{equation}
\label{eqn:jsv=0}
J_s^{(\lambda)} (T) = {\widetilde J}^{(\lambda)} (T)
\times ({\hat \eta_L} 
\cdot 
{\hat \eta_R}) 
\times \sin(\bar \phi),
\end{equation}
where
$
{\widetilde J}^{(\lambda)} (T) = 
(2e / \hbar) 
\Delta_{L} (T) \Delta_{R} (T) S_{LR}^{(\lambda)} (T),
$
$\bar \phi = \phi_R - \phi_L$,
and the label $\lambda$ identifies 
incoherent $( \lambda = inc)$ or
coherent $( \lambda  = coh) $ processes.
Here, 
$S_{LR}^{(\lambda)} (T) = 
\sum_{{\bf k}_L, {\bf k}_R} 
Q^{(\lambda)} ({\bf k}_L, {\bf k}_R)
\psi^{{\widetilde \Gamma}_L, u} ({\bf k}_L)
\psi^{{\widetilde \Gamma}_R, u} ({\bf k}_R)
P({\bf k}_L, {\bf k}_R)$, 
where 
$ P({\bf k}_L, {\bf k}_R)
= 4 T \sum_{m = 0}^{\infty} 
(\hbar^2\nu_m^2 + E_{k_L}^2)^{-1}
(\hbar^2\nu_m^2 + E_{k_R}^2)^{-1}
$, 
and 
$ \psi^{{\widetilde \Gamma}, u} ({\bf k}_L)
= X,~Y,~Z~{\rm or}~XYZ
$.
Notice that 
${\hat \eta}_L \cdot {\hat \eta}_R = \cos(\theta_{LR})$,
where $\cos(\theta_{LR})$ is the angle between the
two ${\bf d}$-vectors, and describes
a polarization angle just like the Malus's law of 
electric polarization. 
$J_s (T)$ 
depends crucially on the 
tunneling matrix elements $Q({\bf k}_L, {\bf k}_R)$.
For purely incoherent processes 
the {\it only} terms that contribute to $J_s^{(inc)} (T)$
come from 
$Q^{(inc)} = 
2 T_s^{1_L 1_R}
T_s^ {{\widetilde \Gamma}_L {\widetilde \Gamma}_R }
- 2T_s^{1_L {\widetilde \Gamma}_R }
T_s^{ {\widetilde \Gamma}_L 1_R }
$, 
where ${\widetilde \Gamma}_j$ are appropriate odd representations 
of the ${\bf d}_j$-vectors. 
For the ``$p_x$'' symmetry 
$
\psi^{{\widetilde \Gamma}_L, u} ({\bf k}_L) = X_L
$, and
$
\psi^{{\widetilde \Gamma}_R, u} ({\bf k}_R) = X_R
$,
thus
(a)
$
Q^{(inc)}_{p_x p_x} ({\bf k}_L, {\bf k}_R) = 
\widetilde Q_{p_x p_x} X_L X_R   
$
produces a non-vanishing $J_s (T)$, 
while 
(b) 
$
Q^{(inc)}_{p_x p_x} ({\bf k}_L, {\bf k}_R) = 
\widetilde Q_{p_x p_x} 1_L 1_R   
$
produces a trivially vanishing $J_s (T)$.
Similarly
for the ``$p_y$'' symmetry 
$
\psi^{{\widetilde \Gamma}_L, u} ({\bf k}_L) = Y_L
$, and
$
\psi^{{\widetilde \Gamma}_R, u} ({\bf k}_R) = Y_R
$,
thus
(c)
$
Q^{(inc)}_{p_y p_y} ({\bf k}_L, {\bf k}_R) = 
\widetilde Q_{p_y p_y} Y_L Y_R   
$
produces a non-vanishing $J_s (T)$, 
while 
(d) 
$
Q^{(inc)}_{p_y p_y} ({\bf k}_L, {\bf k}_R) = 
\widetilde Q_{p_y p_y} 1_L 1_R   
$
produces a trivially vanishing $J_s (T)$.
For comparison purposes, we 
analyse next four coherent tunneling 
processes allowed by group theory corresponding to matrix
elements 
(a) 
$
Q^{(coh)}_{p_x p_x} ({\bf k}_L, {\bf k}_R) = 
\widetilde Q_{p_x p_x} X_L X_R   
\delta_{ {\bf k}_{{\parallel}_L}, {\bf k}_{{\parallel}_R} }   
$,
and
(b) 
$
Q^{(coh)}_{p_x p_x} ({\bf k}_L, {\bf k}_R) = 
\widetilde Q_{p_x p_x} 1_L 1_R 
\delta_{ {\bf k}_{{\parallel}_L}, {\bf k}_{{\parallel}_R} }   
$,
for the ``$p_x$'' symmetry;
(c) 
$
Q^{(coh)}_{p_y p_y} ({\bf k}_L, {\bf k}_R) = 
\widetilde Q_{p_y p_y} Y_L Y_R,   
\delta_{ {\bf k}_{{\parallel}_L}, {\bf k}_{{\parallel}_R} }   
$,
and
(d)
$
Q^{(coh)}_{p_y p_y} ({\bf k}_L, {\bf k}_R) = 
\widetilde Q_{p_y p_y} 1_L 1_R,   
\delta_{ {\bf k}_{{\parallel}_L}, {\bf k}_{{\parallel}_R} }   
$,
for the ``$p_y$'' symmetry,
where the momentum parallel to the junction is conserved.
In Fig.~\ref{fig:one} we show both the temperature 
and angular dependence of $J_s^{(\lambda)}$ for the case of 
${\bf a}$ axis tunneling, assuming that  
${\bf d}_j \parallel {\bf c}_j$ $(j = L,R)$. 
$J_s^{(\lambda)} (T)/J_s^{(\lambda)} (0)$  
is shown for $\alpha_L = 0$, where {\it all}  L and R axes  
coincide. In the inset, we show 
$J_s^{(\lambda)} (0, \alpha_{LR})/J_s^{(\lambda)} (0, 0) $ for $T = 0$.
For coherent processes 
${\bf k}_{{\parallel}_L} = {\bf k}_{{\parallel}_R}$ 
and the parallel momenta are related by 
$k_{y_R} = \cos(\alpha_{LR}) k_{y_L} - \sin(\alpha_{LR}) k_{z_L}$,
and
$k_{z_R} = \sin(\alpha_{LR}) k_{y_L} + \cos(\alpha_{LR}) k_{z_L}$, 
where $\alpha_{LR}$ is the angle between ${\bf k}_{\parallel_L}$
and ${\bf k}_{\parallel_R}$.
For ${\bf a}$ axis tunneling $\theta_{LR} = \alpha_{LR}$.
Notice in the inset of Fig.~{\ref{fig:one}} that the Josephson current
does ${\bf not}$ change sign for the ``$p_y$'' with 
$Q^{(coh)} \propto 1_L 1_R$ (double-dotted line).
This is a direct manifestation of the {\it polarization} effect
of the ${\bf d}$ vector. It is important to emphasize that
both the temperature and the angular dependencies of 
$J_s (T, \alpha_{LR})$ can help distinguish different
symmetries, and 
different (coherent or incoherent) 
tunnelling processes (see Fig.~\ref{fig:one}), 
as seen experimentally for ${\bf c}$ axis tunneling of 
cuprate superconductors~\cite{li-99}.
The approach used for  ${\bf a}$ axis tunneling 
is good only in the cases of ``$p_y$'' or ``$p_z$'' symmetry
but not for ``$p_x$'' or ``$p_{xyz}$'', since the ${\bf d}$-vector
changes sign upon reflection at the interface, and
leads to low energy surface bound states to be discussed next.

{\it The Josephson effect including surface bound states:} 
To obtain $J_s (T)$ for the ``$p_x$'' symmetry 
in the case of ${\bf a}$ axis tunneling 
it is crucial to solve Eqs. (\ref{eqn:bdg1}), (\ref{eqn:bdg2})
and (\ref{eqn:order-parameter}) self-consistently.
We solve the non-local lattice BdG equations and compare
our results with standard local quasiclassical continuum 
approximations~\cite{tanaka-96,barash-96,lofwander-01}.
First, we take advantage of the translational invariance along the 
direction parallel to the interface and write
the BdG amplitudes as
$
u_{N_j \alpha_j} ({\bf r}_j) = 
\exp(i {\bf k}_{j\parallel} \cdot {\bf  r}_{j\parallel})
\tilde u_{{\bf k}_{j\parallel}, {\tilde N}_j, n_j} ({\bf r}_{\perp}) 
/\sqrt{V};
$
$
v_{N_j \alpha_j} ({\bf r}_j) = 
\exp(i {\bf k}_{j\parallel} \cdot{\bf  r}_{j\parallel})
\tilde v_{{\bf k}_{j\parallel}, {\tilde N}_j, n_j} ({\bf r}_{\perp}) 
/\sqrt{V},
$
where quantum indices $N_j$ are represented by momentum 
${\bf k}_j$ and discrete index $n_j = 1_j, 2_j$. 
After a Fourier transformation in the parallel coordinates
the lattice BdG equations ({\ref{eqn:bdg1}) and
(\ref{eqn:bdg2}}) become one-dimensional in
the perpendicular coordinates 
${\bf r}_{\perp}, {\bf r^\prime}_{\perp}$.
We choose hard boundary conditions, where the BdG amplitudes
vanish when ${\bf r}_{{\parallel}_L} = 
{\bf r}_{{\parallel}_R} = 0$ (at the center of the insulating
barrier). 
Second, only the triplet channel component
$V_j^{(t)} ({\bf r}_j, {\bf r}_j^\prime)$
of the general weak spin-orbit coupling interaction
$V_{j, \alpha_j, \beta_j, \delta_j, \gamma_j} ({\bf r}_j, {\bf r^\prime}_j)$ 
is considered.
The triplet component is
$V_j^{(t)} ({\bf r}_j, {\bf r}_j^\prime) = 
I_{1,j} ({\bf r}_j, {\bf r^\prime}_j) +
I_{2,j} ({\bf r}_j, {\bf r^\prime}_j) 
$,
and it is assumed that $V_j^{(t)} ({\bf r}_j, {\bf r}_j^\prime)$ 
has an on-site $({\bf r}_j^\prime = {\bf r}_j)$ 
repulsion $U_j$ and nearest neighbor attractions 
$({\bf r}_j^\prime = {\bf r}_j + {\bf a}_{\eta})$ 
$V_{j, x}$, $V_{j,y}$, and $V_{j,z}$ only. 
In addition, we assume that the direction of the 
${\bf d} ({\bf r}, {\bf r}^\prime)$ is independent of position
and points along the ${\bf c}$ axis.
Furthermore, we choose the spin quantization axis to be
along the direction of ${\bf d} ({\bf r}, {\bf r}^\prime)$. 
Under these conditions the order parameter matrix 
$
\Delta_{j,\alpha_j, \beta_j} ({\bf r}_j, {\bf r^\prime}_j)
$
for the ``$p_x$'' symmetry 
vanishes identically when 
${\bf r^\prime}_j = {\bf r}_j$, 
and has only off-diagonal matrix elements
$
\Delta_{j,\uparrow_j, \downarrow_j} ({\bf r}_j, {\bf r^\prime}_j)
= V_{j,x} 
\left[ 
\langle c_{j,\uparrow_j}({\bf r^\prime}_j) 
c_{j,\downarrow_j} ({\bf r}_j) \rangle
+
\langle c_{j,\downarrow_j}({\bf r^\prime}_j) 
c_{j,\uparrow_j} ({\bf r}_j) \rangle
\right]/2
$
and
$
\Delta_{j,\downarrow_j, \uparrow_j} ({\bf r}_j, {\bf r^\prime}_j)
=
\Delta_{j,\uparrow_j, \downarrow_j} ({\bf r}_j, {\bf r^\prime}_j)
$
when 
${\bf r^\prime}_j = {\bf r}_j + a_j {\bf {\hat a}}$.
$J_s (T)$ is calculated numerically using
Eq. (\ref{eqn:pair-current}) with 
$T_{\alpha_L \alpha_R}  ({\bf r}_L, {\bf r}_R) = T_s ({\bf r}_L, {\bf r}_R)
\delta_{\alpha_L \alpha_R}$.
For definiteness and simplicity we also assume that 
$T_s ({\bf r}_L, {\bf r}_R) = 
\widetilde T \delta_{{\bf r}_{L\parallel}, {\bf r}_{R\parallel}}
$
for $r_{L\perp} = -d/2 $ 
and $r_{R\perp} = +d/2 $;
$T_s ({\bf r}_L, {\bf r}_R) = 0$ otherwise. 
Here, $d$ is the separation between the last layer
of the L and the first layer of the R superconductor, 
and corresponds to the {\it thickness} of the insulating layer.
This simple choice guarantees that the parallel momentum
${\bf k}_{\parallel}$ is conserved across the junction, and thus
correspond to a coherent process.
Since the tunneling matrix elements connect mostly 
states that have appreciably large BdG
amplitudes near the junction,
we use a mixed representation of Eq.~(\ref{eqn:pair-current}) expressed
in terms of the parallel momenta 
${\bf k}_{{\parallel}_L} = {\bf k}_{{\parallel}_R} = {\bf k}_{\parallel}$ 
and the perpendicular coordinates
${\bf r}_{{\perp}_L}$ and ${\bf r}_{{\perp}_R}$
to calculate numerically $J_s (T)$.
In Fig.~{\ref{fig:two} we compare our results for $J_s (T)$
using the non-local lattice BdG equations and the local quasiclassical
approaches~\cite{tanaka-96,barash-96,lofwander-01}. 
\begin{figure} [hb!]
\begin{center}
\epsfxsize=8cm
\epsfysize=8cm
\epsffile{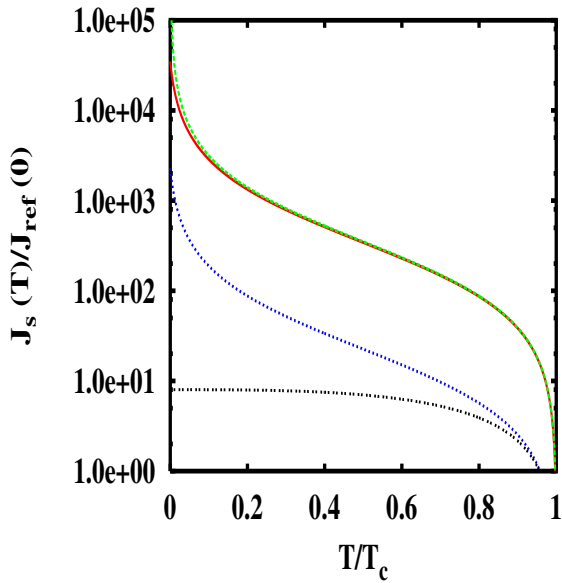}
\vskip 0.3cm
\caption{
Plots of $J_s (T)/J_{\rm ref} (0)$ versus $T/T_c$ for 
for coherent ``$p_x$'' (``$p_z$'') ${\bf a}$ (${\bf c}$) 
axis tunneling, where
$J_{\rm ref} (0)$ is the Landau critical current
for a singlet s-wave superconductor with the same $T_c$.
Quasiclassical results are indicated by dashed (dotted) lines
for ``$p_x$'' (``$p_z$''), while the non-local BdG results
are represented by solid (double-dotted) lines
for ``$p_x$'' (``$p_z$''), for the parameters of Fig~\ref{fig:one}.
}
\label{fig:two}
\end{center}
\end{figure}
The local quasiclassical approximation is only strictly valid
when $k_{F_\perp} \xi_{\perp} \to \infty$, but it leads 
to {\it zero} energy bound states in the case where 
the quasiclassical order parameter $\Delta ({\bf k}, {\bf r})$
changes 
sign upon ${\bf r} \to -{\bf r}$~\cite{tanaka-96,barash-96,lofwander-01} 
(as in the case of the ${\bf a}$ axis tunneling for the ``$p_x$'' symmetry). 
However, in the non-local lattice BdG equations (\ref{eqn:bdg1})
and (\ref{eqn:bdg2}) there are 
only {\it finite} energy bound states.
The non-existence of {\it zero} energy bound
states in the non-local lattice BdG equations can be understood
as follows. Because the order parameter is non-local, it can be
described in terms of the center of mass ${\bf R} = [{\bf r} +
{\bf r^\prime}]/2$ and relative ${\bf r}_{rel} = 
[{\bf r^\prime} - {\bf r}]$ coordinates. 
Near the surface these two coordinates are entangled and 
thus lift the {\it zero} energy bound states degeneracy found in the
local theory. This produces, perturbatively, finite energy 
bound states of the order $|\Delta_0|/\gamma_{\perp}$ where
$\gamma_{\perp} = \xi_{\perp}/a_{\perp}$, with 
$\xi_{\perp}$ and $a_{\perp}$ being the coherence and 
unit cell lengths, respectively.
The finiteness of the energy of the bound states
cuts off the low temperature $1/T$ divergence 
of $J_s (T)$ calculated in the quasiclassical approach 
at small transparencies~\cite{tanaka-96,barash-96,lofwander-01}. 
For ${\bf a}$ axis tunneling and the ``$p_x$'' symmetry this amounts
to a small correction to the quasiclassical results, as
$\gamma_{x} = \xi_{x}/a_{x} \approx 106$ for the Bechgaard salts 
and $\Delta_{0,p_x} = 3.73~{\rm K}$, leading to the lowest bound
state energy to be approximately $ T_{p_x}^* = 35.2~{\rm mK}$. 
Which means that for $T < T_{p_x}^*$ the quasiclassical results
are not reliable (see Fig.~\ref{fig:two}). 
The quasiclassical results fail in a more dramatic way, when 
we consider ${\bf c}$ axis tunneling for the ``$p_z$'' symmetry. 
In this case $\xi_{z} = 21.5~{\rm \AA}$, $a_{z} = 13.5~{\rm \AA}$,
$\gamma_{z} = \xi_{z}/a_{z} \approx 1.59$ and 
$\Delta_{0,p_z} =  3.21~{\rm K}$, 
resulting in $T_{p_z}^* = 2.02~{\rm K}$, which is larger than
the critical temperature $T_c = 1.5~{\rm K}$ used here. 
Thus, perturbative corrections to quasiclassical {\it zero} energy bound
states are enormous, and the quasiclassical approximation can not
be used except for $T \approx T_c$ 
(see Fig.~\ref{fig:two}).

In summary, we have discussed the Josephson current $J_s$ in 
TS/I/TS junctions, and indicated that the triplet nature of the
order parameter for Bechgaard salts could be tested in 
such an experiment. 
We emphasized that the temperature and angular dependence 
of $J_s$ can help distinguish between different triplet
order parameter symmetries, and we discussed the role of 
surface bound states within the non-local lattice BdG equations.
We would like to thank NSF (Grant No. DMR-980311)
for financial support.

\end{document}